\documentclass[letterpaper,journal]{IEEEtran}
\usepackage{amsmath,amsfonts}
\usepackage{algorithmic}
\usepackage{siunitx}
\usepackage{bm}
\usepackage{booktabs}
\usepackage{hyperref}
\usepackage{cleveref}

\usepackage{graphicx} % DO NOT CHANGE THIS
\usepackage{array}
\DeclareSIUnit\angstrom{\text{Å}}
\usepackage[caption=false,font=normalsize,labelfont=sf,textfont=sf]{subfig}
\usepackage{textcomp}
\usepackage{stfloats}
\usepackage{url}
\usepackage[numbers]{natbib} 
\usepackage{verbatim}
\usepackage{graphicx}
\def\BibTeX{{\rm B\kern-.05em{\sc i\kern-.025em b}\kern-.08em
    T\kern-.1667em\lower.7ex\hbox{E}\kern-.125emX}}
\usepackage{balance}

\begin{document}

\title{MIN: Multi-channel Interaction Network for \\ Drug-Target Interaction with Protein Distillation}

\author{
    Shuqi Li\textsuperscript{*}, Shufang Xie\textsuperscript{*}, Hongda Sun\textsuperscript{*}, Yuhan Chen, Tao Qin, Tianjun Ke, Rui Yan% <- this stops a space
    \IEEEcompsocitemizethanks{
        \IEEEcompsocthanksitem Shuqi Li, Shufang Xie, Hongda Sun, Yuhan Chen, Tianjun Ke, and Rui Yan are with the Gaoling School of Artificial Intelligence, Renmin University of China, Beijing, China, 100080.
        E-mail: \{shuqili, shufangxie, sunhongda98, yuhanchen, keanson, ruiyan\}@ruc.edu.cn
        \IEEEcompsocthanksitem Tao Qin is with Microsoft Research.
        E-mail: taoqin@microsoft.com
        \IEEEcompsocthanksitem Corresponding author: Rui Yan (ruiyan@ruc.edu.cn)
        \IEEEcompsocthanksitem * Equal Contribution.
    }
}

\markboth{Journal of \LaTeX\ Class Files,~Vol.~18, No.~9, September~2020}%
{Li \MakeLowercase{\textit{et al.}}: MIN: Multi-channel Interaction Network}

\maketitle
\begin{abstract}

Traditional drug discovery processes are both time-consuming and require extensive professional expertise. With the accumulation of drug-target interaction (DTI) data from experimental studies, leveraging modern machine-learning techniques to discern patterns between drugs and target proteins has become increasingly feasible. In this paper, we introduce the Multi-channel Interaction Network (MIN), a novel framework designed to predict DTIs through two primary components: a representation learning module and a multi-channel interaction module. The representation learning module features a C-Score Predictor-assisted screening mechanism, which selects critical residues to enhance prediction accuracy and reduce noise. The multi-channel interaction module incorporates a structure-agnostic channel, a structure-aware channel, and an extended-mixture channel, facilitating the identification of interaction patterns at various levels for optimal complementarity. Additionally, contrastive learning is utilized to harmonize the representations of diverse data types. Our experimental evaluations on public datasets demonstrate that MIN surpasses other strong DTI prediction methods. Furthermore, the case study reveals a high overlap between the residues selected by the C-Score Predictor and those in actual binding pockets, underscoring MIN's explainability capability. These findings affirm that MIN is not only a potent tool for DTI prediction but also offers fresh insights into the prediction of protein binding sites.
\end{abstract}


\section{Introduction}
\label{intro}

Drug-target Interaction (DTI) prediction is vital in drug re-purposing and drug discovery. Although biological assays such as high-throughput screening remain the mainstream approach to identifying DTI, they are generally inefficient in practical applications since they are expensive and time-consuming~\cite{haggarty2003multidimensional, chen2016drug}. Thus, utilizing computational methods to predict potential interactions with reduced time and less experimental efforts from massive data is of great significance in drug discovery.

\begin{figure}
  \centering
\includegraphics[width=\linewidth]{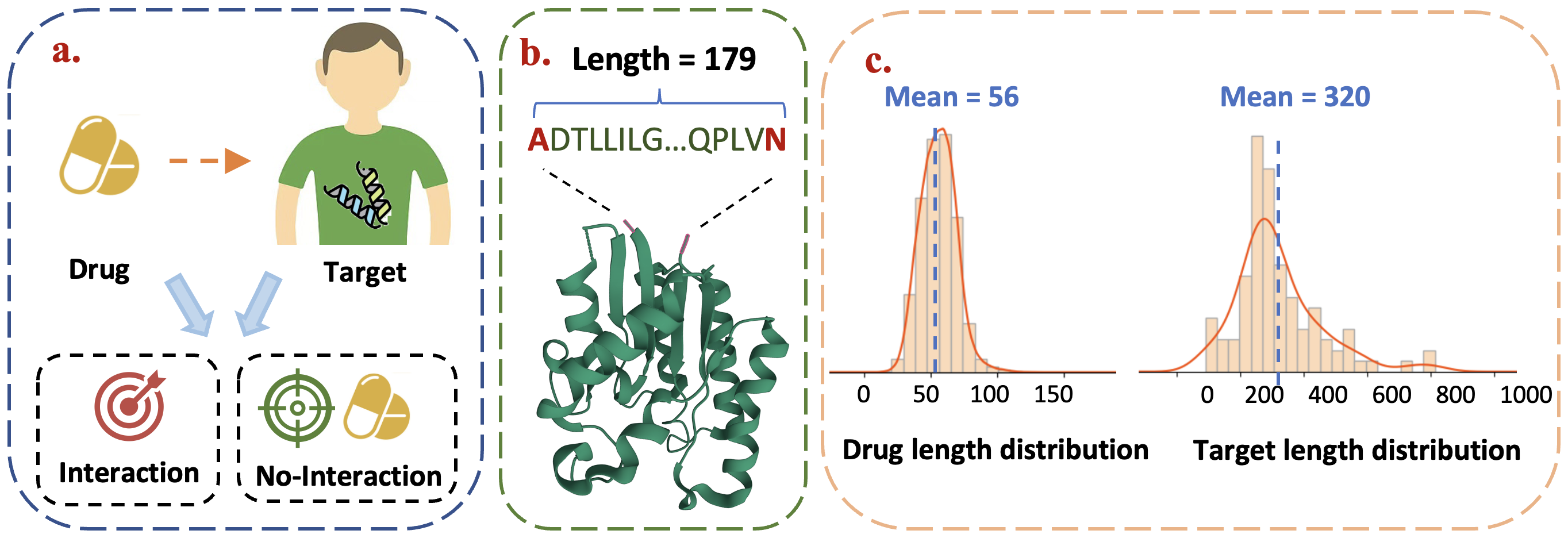}
  \caption{(a) A sketch map of the drug-target interaction classification task. (b) An example illustrates that residues can be far away in the sequence but are spatially close to each other after folding. (c) The average length of proteins is much longer than that of drugs (the results are computed on the DUD-E dataset).}
  \label{fig:task}
\end{figure}

Over the past decade, numerous computational methods have been developed to predict drug-target interactions (DTIs). Traditional similarity-based approaches, which utilize the similarity between drugs and target proteins, often yield suboptimal predictions when only a few known binding drugs are associated with a target~\cite{keiser2007relating, keiser2009predicting}. Additionally, molecular docking methods have been used to screen interaction pairs by analyzing the spatial structures of drug compounds and target proteins~\cite{cheng2007structure,donald2011algorithms,fu2018predictive}. These methods typically depend on identifying binding residues within the binding pocket by estimating the binding score. However, the broad application of molecular docking is hindered by time constraints. With the accumulation of extensive DTI data, we can now utilize machine learning to extract matching patterns for prediction. The formulation of the DTI task varies, including recommendation~\cite{alaimo2016recommendation}, regression (binding affinity prediction), supervised classification~\cite{wen2017deep}, and link prediction in bipartite graphs~\cite{Bleakley2009Supervised}. In this paper, we model DTI prediction as a classification task, as illustrated in Figure~\ref{fig:task} (a), which involves determining whether a given drug and target interact.

Most previous research focused on modeling drugs and proteins as sequences because they are simpler to formulate and compute~\cite{weininger1988smiles}. However, the modest prediction performance of these methods suggests that sequence data alone may not provide sufficient information. With advances in structural biology, the availability of protein structures has increased, enhancing the DTI process with crucial spatial information, such as cavities and surfaces. This allows for multimodal modeling of DTI tasks. Recent efforts aim to integrate structural information to improve DTI prediction accuracy~\cite{ragoza2017protein,stepniewska2018development}. Some of them attempt to serve characterized structure as input,
such as graph representation for drug molecules ~\cite{torng2019graph} and distance map for the target proteins~\cite{zheng2020predicting}. 

Although a variety of data sources are available, effectively aligning them remains a significant challenge. This is because residues that are distant in a sequence can be spatially close after folding, as depicted in Figure~\ref{fig:task} (b). Another issue is the substantial length disparity between target proteins and drug compounds. For instance, in the DUD-E dataset, we have analyzed and visualized the length distributions of target proteins and drug compounds in Figure~\ref{fig:task} (c), where the average length of proteins is $320$, compared to just $56$ for drugs. This considerable difference complicates the matching of drugs to critical residues and locating the binding pocket, which is crucial for accurate DTI prediction.

To tackle these issues, we propose a novel DTI prediction framework, called Multi-channel Interaction Network (MIN). Our main contributions can be summarized as follows:
\begin{itemize}

\item Multi-channel Interaction Network (MIN) unifies both the sequence and structure information of target proteins and drug compounds. For better information alignment, we design three interaction channels (structure-agnostic, structure-aware, and extended-mixture channels) to extract interaction patterns at different levels for mutual complementation.

\item A C-Score Predictor is proposed that dynamically distills (i.e. filters out unimportant residues) protein sequences using biological conservation information to improve prediction efficiency and reduce noise for better locating potential binding pockets. 
\end{itemize}

We conduct extensive experiments on two widely used benchmarks, the DUD-E datasets, and Human datasets, and achieve state-of-the-art performance in various evaluation metrics. We also implement detailed case studies to present the explainability of MIN.

\section{Related Work}
\begin{figure*}
  \centering
  \includegraphics[width=18cm]{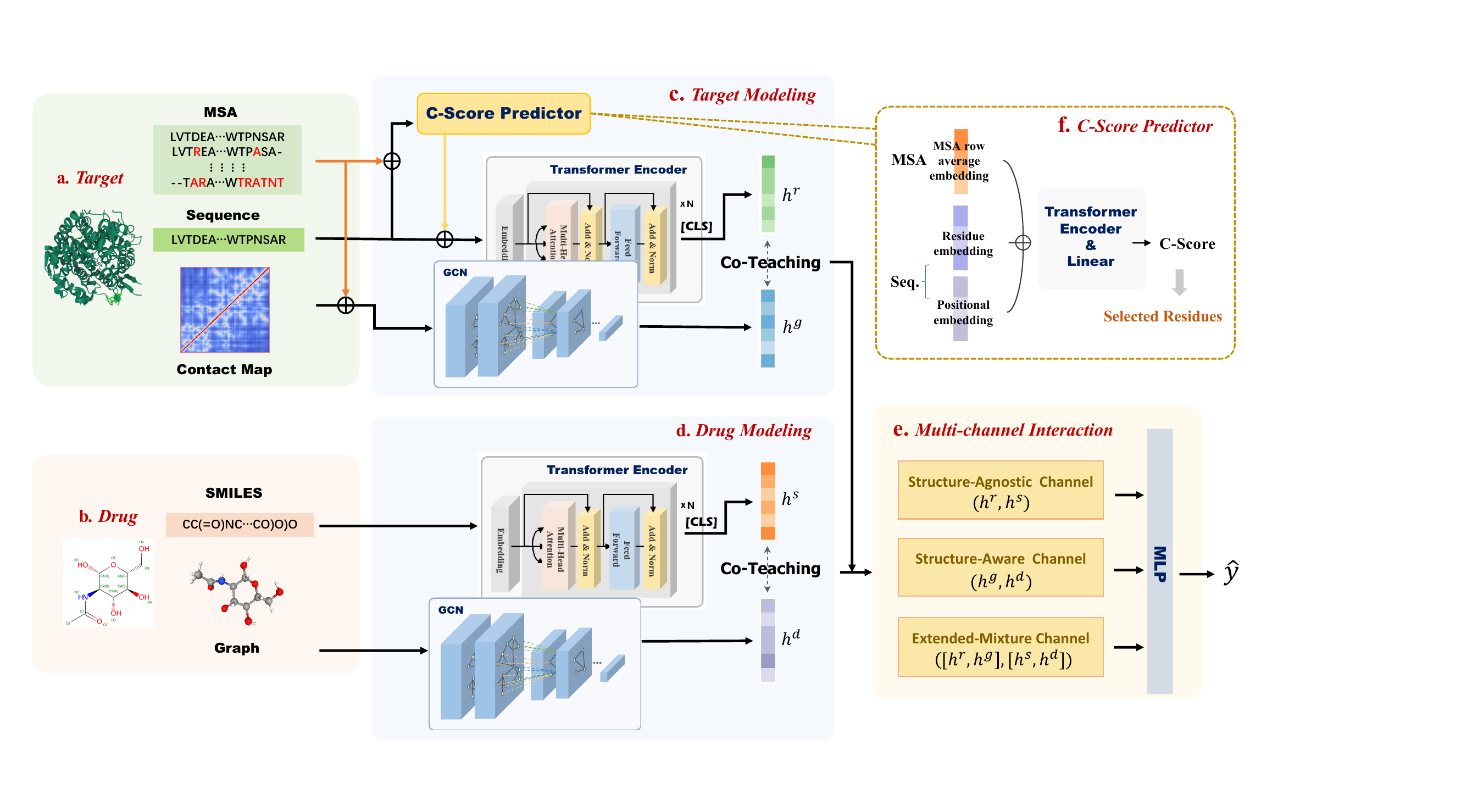}
  \caption{An overview of MIN; (a) Target protein inputs; (b) Drug inputs; (c) Target protein representation learning; (d) Drug representation learning; (e) Multi-channel interaction; (f) C-Score Predictor.}
  \label{fig:framework}
\end{figure*}

With the increasing development of deep learning, deep neural models are used to predict DTIs due to their powerful representation ability. In this paper, the related work covered is limited only to deep learning approaches.

DeepDTA~\cite{ozturk2018deepdta} and DeepAffinity~\cite{karimi2019deepaffinity} were representatives of deep-learning-based DTI models that require only SMILES strings of drugs and amino acids sequences of proteins to mine deep representations of input sequences.~\citet{2020DeepPurpose} proposed DeepPurpose, a comprehensive and easy-to-use deep learning library for DTI prediction. ~\citet{li2020monn} developed a multi-objective network using GCN~\cite{kipf2016semi} and CNN to process input features without structure information. ~\citet{2011NNScore} presented a scoring function based on a multi-layer perception neural network to characterize the binding affinities of protein-ligand complexes. ~\citet{zheng2020predicting} predicted DTIs by representing proteins with a two-dimensional distance map from monomer structures. ~\citet{cai2021msa} developed DISAE for the protein sequence representation, which can utilize all protein sequences and their Multiple Sequences Alignments (MSA) to capture functional relationships between proteins without the knowledge of their structure and function. ~\citet{chen2020transformercpi} designed a rigorous label reversal experiment to test whether a model learns true interaction features to make a better prediction.~\citet{WANG2021107476} employed the compound graph and protein sequence data for both interaction classification and binding affinity prediction tasks. ~\citet{wu2022bridgedpi} introduced hyper-node to bridge different proteins/drugs to work as protein-protein associations and drug-drug associations. AttentionSiteDTI (ASDTI)~\cite{yazdani2022attentionsitedti} utilizes protein binding sites along with a self-attention mechanism to address the problem of drug-target interaction prediction.
~\citet{moon2022pignet} designed a novel physics-informed graph neural network named PIGNet, which considered the four energy components – van der Waals (vdW) interaction, hydrogen bond, metal-ligand interaction, and hydrophobic interaction to predict the binding affinity.~\citet{SONG2022269} proposed DeepFusion, which generated global structural similarity feature and generate local chemical sub-structure semantic feature respectively for both drug and protein.~\citet{dehghan2023tripletmultidti} proposed TripletMultiDTI, which focuses on multimodal representation learning in DTI prediction using a triplet loss function. ~\citet{feng2024gcardti} introduced GCARDTI, a hybrid mechanism framework based on convolutional neural network and graph attention network for capturing multi-view feature information of drug and target molecular structures.~\citet{liu2024higraphdti} proposed HiGraphDTI, a hierarchical graph representation learning method for DTI prediction to overcome limitations in sequential model-based target feature extraction. ~\citet{zhao2024pocketdta} designed PocketDTA, an advanced drug target affinity prediction model that enhances generalization through the pre-trained sequence and 3D structure representations, combines global and local multimodal data with fewer parameters, and uses a bilinear attention network for interpretable drug-target interaction analysis. ~\citet{gao2024drugclip} proposed DrugCLIP, by reformulating virtual screening as a dense retrieval task and employing contrastive learning to align representations of binding protein pockets and molecules from a large quantity of pairwise data without explicit binding-affinity scores. ~\citet{svensson2024hyperpcm} introduced HyperPCM, which adopt a HyperNetwork-specific
initialization strategy to ensure prediction stability and enrich the protein target representations using a context module.

In this paper, we take advantage of conservation information and design a novel multi-channel interaction network, which learns interaction patterns between drugs and targets at different levels.

\section{Preliminaries and Formulation}

\subsection{Preliminaries}
\label{Preliminaries} 

Before diving into the details of the MIN model, we first introduce some necessary concepts, which are the basis of this paper.

\paragraph{Target Protein Sequence}  A target protein with length $L$ can be represented as a sequence of discrete residue characters $ (r_1, r_2,\cdots, r_L) \in R^{1 \times L}$ in the standard 20-character alphabet, which could further encoded as an embedding matrix $P_r \in R^{h \times L}$ with hidden size $h$.

\paragraph{Multiple Sequences Alignment (MSA)}  MSA is a sequences alignment matrix of multiple homologous protein sequences for a target protein. Given a protein sequence, the MSA could be generated by searching the similar sequences by performing pairwise comparisons from a database~\footnote{In this study, we use the protein database named Uniclust 30~\cite{10.1093/nar/gkw1081}.}. Here we formulate an MSA matrix as $M \in R^{m \times L}$, where $m$ indicates the number of searched sequences and $L$ is the length of the given protein sequence.

\paragraph{Position Specific Scoring Matrix (PSSM)} PSSM is a commonly used representation of motifs in protein sequences. Based on a protein's MSA matrix, this protein's PSSM $P_p \in R^{20 \times L}$ could be estimated. Every column in PSSM is a distribution denoting the occurrence possibility of every residue in the standard 20-character alphabet.

\paragraph{Conservation Score (C-Score)} For a protein with length $L$, C-Score $y_s \in R^{1\times L}$ quantifies every residue conservation in a given MSA by calculating the degree of residue variability in each column. Highly conservative residues are presumed to be functionally or structurally important because they have accepted fewer mutations relative to the rest of the alignments~\cite{msa}.

\paragraph{Protein Contact Map} A contact map $P_{e}\in R^{L\times L}$ is a binary two-dimensional pairwise matrix, which denotes whether amino acid pairs are "in contact" or not in 3D space. Each entry in $P_{e}$ equals to $1$ if $ 1/(1+d / d_{0})\leq s_0$, where $d$ is the distance between $C\alpha$ atoms of  the residue pair. $d_0 = 3.8 \qty{4}{\angstrom}  $ is a constant to normalize contact distance, and $s_0$ the contact threshold~\cite{zheng2020predicting}.

\paragraph{The Simplified Molecular-Input Line-Entry System (SMILES)} SMILES $D_s\in R^{1\times n}$ is a specification in the form of a line notation for representing the molecules using short ASCII strings, where $n$ is the length of a molecule SMILES. For example, the benzene ring is represent as 'c1ccccc1' in SMILES format.

\subsection{Formulation}
We formulate DTI prediction as a binary classification task with taking drug and target as inputs.

\paragraph{Target Inputs.} A target protein can be represented by a residue sequence representation $P_r$. We also design the MSA-guided contact graph $P_g = \left\{P_u, P_{e} \right\}$ to capture the protein's structure information in 3D space. In $P_g$, residue nodes $P_u$ are denoted as PSSM $P_p$ weighted residue embeddings. Specifically, the first residue node feature is calculated from all $20$ residue embeddings weighted by the first column of $P_p$. This kind of node representation considers all the potential residues at the same position so that it contains information provided by MSA. Besides, the contact map $P_e$ would serve as the adjacency matrix.

\paragraph{Drug Inputs.} A drug sequence can be simply described as a SMILES $D_s\in R^{1\times n}$. Besides, graph format $D_g = \{D_v, D_e\}$ is used to express drug structure, where each node in $D_v$ corresponds to a non-hydrogen atom, and each edge in $D_e$ represents a chemical bond existing or not.

\paragraph{Problem Formulation.} Our model aims to learn a mapping that takes $(P_r; P_g; D_s; D_g)$ as input and outputs a prediction value $\hat{y} \in {0,1}$, where $\hat{y}=1$ indicates the presence of an interaction between the specified drug-target pair.

\section{Multi-channel Interaction Network}

The overall framework of MIN comprises two major components: representation learning and interaction prediction, which can be further divided into four parts: {\it Target protein representation learning, Drug compound representation learning, Representations contrastive learning}, and {\it Multi-channel interaction prediction}.

\begin{enumerate}
    \item In {\it Target protein representation learning}, we design the C-Score Predictor to distill protein sequences by filtering out those unimportant to DTI task and reveal only the critical residues. Then the distilled protein sequences and MSA-guided contact graph are encoded by the Transformer~\cite{vaswani2017attention} encoder and GCN~\cite{kipf2016semi}, respectively (Figure~\ref{fig:framework}(c)).
 
    \item In {\it Drug compound representation learning}, we leverage the Transformer encoder and GCN to process SMILES and graph, respectively (Figure~\ref{fig:framework}(d)).
 
    \item To align the different level representations, {\it Representations contrastive learning} is applied to restrict representations of the same entity to be similar in the latent space.
 
    \item To enhance interaction performance, we design a {\it Multi-channel interaction network} (Figure~\ref{fig:framework}(e)) to unify three channel information, i.e., structure-agnostic channel, structure-aware channel, and extended-mixture channel.
\end{enumerate}

\subsection{Target Representation Learning.}
\label{target}
To represent a target protein, we first distill the original protein sequence to a shorter one by C-Score Predictor, then use the Transformer encoder and GCN to encode the distilled sequence and the MSA-guided contact graph. Transformer is able to extract sequence patterns, while GCN can capture atom relationships on the graph.

A target's binding pocket is a specific region that interacts with another molecule. Identifying the actual binding pocket can simplify the extraction of drug-target interaction patterns. However, automatically recognizing these pockets remains challenging due to the binding mode's flexibility and the significant size disparity between targets and drugs.  Conservation scores are a good indicator for finding the binding pocket of a protein. Binding pockets are often involved in essential biological processes, such as ligand binding or catalysis. The residues in these pockets play crucial roles and are therefore subject to evolutionary pressure. As a result, they tend to be more conserved across different species to maintain the protein's function. High conservation scores reflect this evolutionary constraint, making them effective indicators for identifying potential binding sites~\cite{guharoy2005conservation}. Besides, binding pockets often have structural features that are essential for their interaction with ligands. The residues forming these structural elements are conserved to maintain the 3D conformation needed for effective ligand binding. Conservation scores can help identify these structurally significant regions within the protein. 

\begin{figure}
  \centering
  \includegraphics[width=7cm]{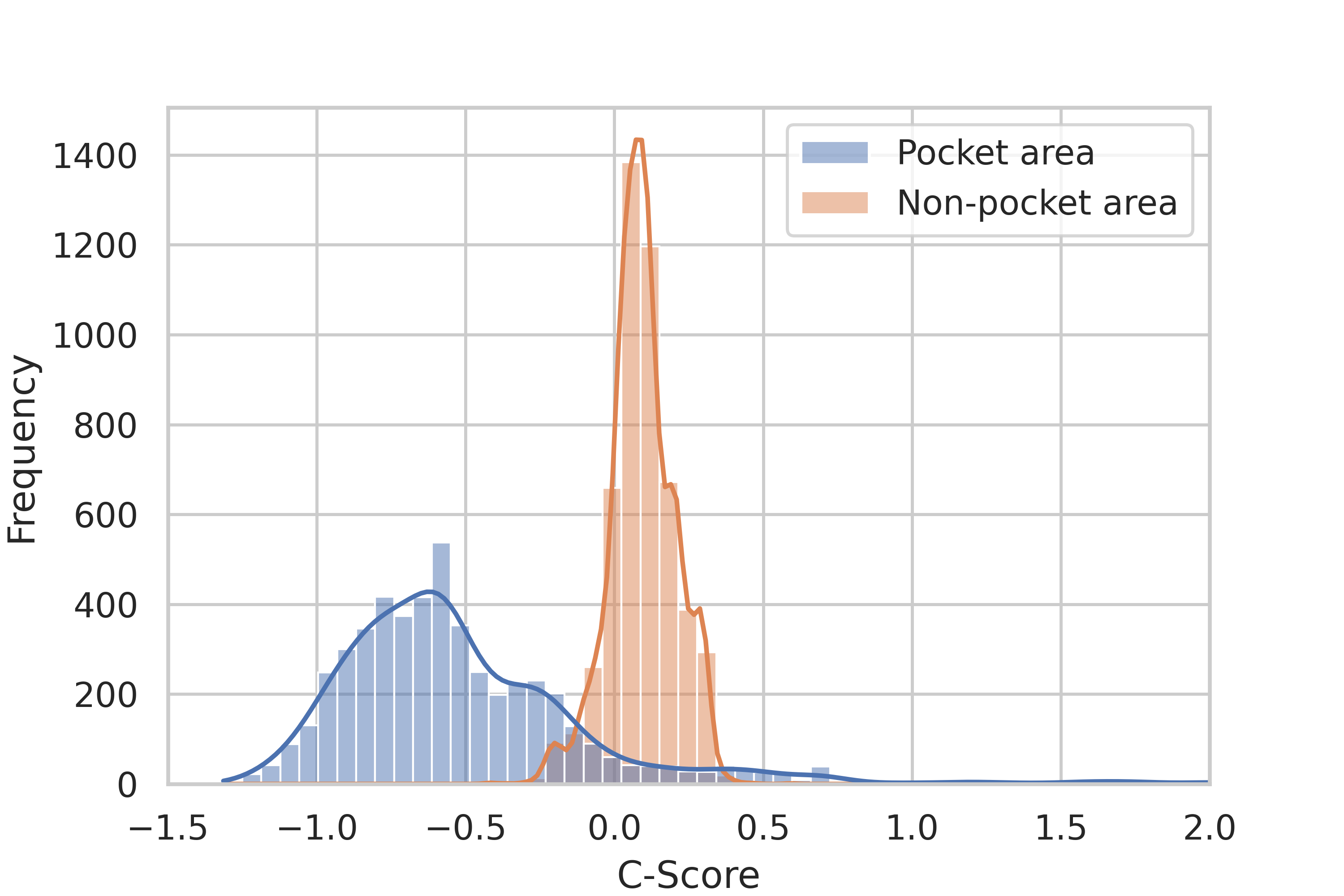}
  \caption{The conservation score (C-Score) difference between the residues in the binding pocket area and non-pocket area. A smaller C-score indicates higher conservation.}

  \label{fig:conservation}
\end{figure}

Assuming that residues in binding pockets are more conserved than those in non-pocket areas, we used the PDBbind v2020 refined set~\cite{wang2004pdbbind} to investigate the conservation differences between these two areas. This dataset contains 5,316 high-quality structures of protein-ligand complexes and natural binding pockets. Using the ConSurf tool~\cite{ben2020consurf, goldenberg2009consurf}, we calculated the mean C-Score for both binding and non-pocket areas for each protein. A lower C-Score indicates higher conservation. The plotted distributions in Figure~\ref{fig:conservation} demonstrate that, on average, residues in the pocket are more conserved than those outside it. Additionally, the P-value from the t-test is significantly less than $0.01$, confirming a statistically significant conservation difference between the two areas. These findings support the value of C-Score for detecting binding pockets in DTI prediction tasks.

Building on these findings, we developed the C-Score Predictor to dynamically retain important residues, thus enhancing prediction accuracy and minimizing noise. Importantly, the C-Score Predictor does not simply use the C-Score for direct cutoffs. Instead, it is also trained with the DTI objective, balancing between conservation and the DTI task label because while conservation is crucial, it is not the sole determinant.

The proposed C-Score Predictor is shown in Figure~\ref{fig:framework} (f), which consists of a Transformer encoder $\varphi_c(\cdot)$ and a fully connected layer $f_c(\cdot)$ to predict C-Score $y_s \in R^{1\times L}$ for every residue. Sequence $P_r$ and MSA matrix $M$ are treated as inputs. Here MSA is used to provide extra information to fit the conservative score better. We use three embedding layers to encode $M$, $P_r$ and position information. 
The $e_M, e_r, e_p$ denote the row embeddings average of each position in $M$, residue embedding in $P_r$, and the positional embedding for each position, respectively.
The MSE optimization loss is used to guide residue selection:
$$\hat{y}_s=f_c(\varphi_c([e_M, e_r, e_p]) )$$
\begin{equation}
\mathcal{L}_{\text{MSE}}(y_s, \hat{y_s})=\sum_{i=1}^{\text{L}}\left(y_s^{i} -\hat{y_s}^i\right)^2
\end{equation}
where the label $y_s \in R^{1\times L}$ is calculated C-Score as mentioned in Preliminaries. 

Then we distill the sequence embedding by masking the residues with low C-Score by a special embedding $e_\text{[MASK]}$. The distilled sequence embedding $\hat{P}_r = [\hat{r}_1, \hat{r}_2, \cdots, \hat{r}_i, \cdots]$ is defined as:
\begin{equation}
\hat{r}_{i}=\left\{\begin{array}{ll}
[e_M, e_r, e_p]_i\; &, \;\hat{y}_s^i\leq r_{0} \\
e_\text{[MASK]}\;\;\;\; &, \;\hat{y}_s^i> r_{0} \\
\end{array}\right.
\end{equation}
where $r_0$ is the given C-Score threshold. Then embedding matrix $\hat{P}_r$ is fed into another transformer encoder $h^r = \varphi_r(\hat{P}_r)$ to extract sequence features of target protein. We add a special token $\texttt{[CLS]}$ to the beginning of the sequence~~\cite{zhu2021dual}, and treat the output hidden vector of $\texttt{[CLS]}$ as the sequence level representation. 

For target structure representation, the MSA-guided contact graph $P_g = \left\{P_u, P_{e} \right\}$ is fed into a GCN $\psi_g(\cdot)$ to obtain residues' features as $h^g = \psi_g(P_g)$, where we use mean readout to aggregate the graph level representation.

It is noteworthy that the MSA-guided contact graph is not distilled to maintain the overall structure of target protein.

\subsection{Drug Representation Learning.} 

For drug representation, we use a Transformer encoder  $\varphi_s(\cdot)$ and a GCN $\psi_d(\cdot)$ to encode the SMILES and drug graph.
$$h^s = \varphi_s(D_s), \;h^d = \psi_d(D_g)$$ 
The use of $\varphi_s(\cdot)$ and $\psi_d(\cdot)$ are similar to $\varphi_r(\cdot)$ and $\psi_g(\cdot)$. 
The vital difference is that no distillation is required because the molecules are relatively small.

\subsection{Representations Contrastive Learning.}

Contrastive learning is an effective method for aligning the representations of drugs and targets because it explicitly trains the model to recognize and minimize the distance between related representations in the latent space while maximizing the distance between unrelated ones~\cite{jiang2023adaptive, xiao2024simple}. This approach is particularly beneficial when the sequence and 3D structural patterns of the drugs and targets are captured separately, necessitating their unification in a common latent space for effective interaction modeling. The primary goal of using contrastive learning in this context is to ensure that the representations of the same entity (be it a drug or a target) become more similar in the latent space, irrespective of whether they are derived from sequence data or structural data.

More specifically, we fist employ four projection modules $\rho_{r}, \rho_{g},$ $\rho_{s}, \rho_{d}$ to project the representations of drug, $h^s$, $h^d$, and target, $h^r$, $h^g$, which are denoted as,
\begin{align}
    p_r &= \rho_{r}(h^r), p_g = \rho_{g}(h^g)\\
    p_s &= \rho_{s}(h^s), p_d = \rho_{d}(h^d).
\end{align}
Each $\rho(\cdot)$ is a 3-layer MLP that projects sequence and structure representations into a latent space where they can be directly compared. 

Next, representations $p_r$ and $p_g$ from the same target are treated as positive pairs, implying that they should be close in the latent space. Conversely, representations from different targets or different types (sequence vs structure) are treated as negative pairs, meaning the model learns to distance them in the latent space.
For example, we treat the representation $p_r$ and $p_g$ from the same target as $(p_r^+,p_g^+)$ as a positive sample and use all other targets in the mini-batch to construct negative samples.
Suppose there are $B$ targets in a mini-batch, we will construct $B-1$ $(p_r^+,p_g^{-})$ pairs and $B-1$ $(p_r^-,p_g^+)$ pairs.
Finally, we employ InfoNCE~\cite{he2020momentum} loss to optimize the representation with the similarity function $\text{sim}(\boldsymbol{u},\boldsymbol{v})=\boldsymbol{u}^{\top} \boldsymbol{v} /\|\boldsymbol{u}\|\|\boldsymbol{v}\|$.

The loss for every target in the training set is defined as:
$$
L=\sum_{i}\exp \left(\text{sim}(p_r^+,p_g^i)/ \tau\right)+\sum_{j}\exp \left(\text{sim}(p_r^j,p_g^+)/ \tau\right)
$$
\begin{equation}
\mathcal{L}_{\text{CLT}}(p_r, p_g)=-\log \frac{\exp \left(\text{sim}(p_r^+,p_g^+) / \tau\right)}{L+\exp \left(\text{sim}(p_r^+,p_g^+) / \tau\right)}
\end{equation}
where $\tau$ denotes the temperature, $i, j \in \{1, 2, \cdots, B-1\}$, and the denominator is over one positive and $2B-2$ negative samples.

The contrastive learning for drug ($\mathcal{L}_{\text{CLD}}$) has the same form as the target. Contrastive learning not only enhances the consistency between different types of representations for the same entities but also improves the model’s ability to generalize from the training data to unseen examples, thereby enhancing its predictive performance in drug-target interaction tasks. 

\subsection{Multi-channel Interaction Network.} We design a tree-channel interaction network to enhance DTI prediction (Figure~\ref{fig:framework}(e)). The details of each channel are as follows.

In the structure-agnostic channel, the drug sequence representation and target sequence representation, i.e., $h_s$ and $h_r$, are treated as inputs of a 3-layer MLP network  $\zeta_{\text{agnostic}}(\cdot)$. This channel aims to capture sequence-level interaction patterns. In the structure-aware channel, we put the two structure representations of drug and target, i.e.,  $h_d$ and $h_g$, into a 3-layer MLP network $\zeta_{\text{aware}}(\cdot)$. This channel is used to extract structure-level interaction patterns. In the extended-mixture channel, we first concatenate the sequence representations and the structure representations to obtain the target mixture representation $h^{m_1} = [h^r, h^g]$ and the drug mixture representation $h^{m_2} = [h^s, h^d]$. Then another 3-layer MLP network $\zeta_{\text{mixed}}(\cdot)$ is leveraged to identify the interaction pattern based on both sequence and structure levels.

After multi-channel interaction, we obtain three representations:
\begin{align}
    c_{\text{agnostic}} &=\zeta_{\text{agnostic}}([h_s, h_r]) \\
    c_{\text{aware}}&=\zeta_{\text{aware}}([h_d, h_g]) \\
    c_{\text{mixed}}&=\zeta_{\text{mixed}}([h^{m_1}, h^{m_2}])
\end{align}

where $[\cdot, \cdot]$ denotes the concatenate operator. The last layer of interaction network is to aggregate the $c_{\text{agnostic}}$, $c_{\text{aware}}$ and $c_{\text{mixed}}$, which is defined as:
\begin{equation}
\hat{y}=\text{sigmoid} (f([c_{\text{agnostic}}, c_{\text{aware}}, c_{\text{mixed}}]))
\label{logit}
\end{equation}
where $f$ indicates a feed-forward layer. Here $\hat{y}$ is the prediction probability of drug-target interaction, and $y$ is the binary label. The prediction objective is to minimize the cross entropy loss $\mathcal{L}_{\text{CE}}$ as follows:
\begin{equation}
\mathcal{L}_{\text{CE}}(y, \hat{y})=-\sum_{i=1}^{\text{n}}\left(y_{i} \log \left(\hat{y}_i\right)+\left(1-y_{i}\right) \log \left(1-\hat{y}_i\right)\right)
\end{equation}
where $n$ is the total number of DTI pairs in the training set.

\noindent\textbf{Learning Objective.}
Our model MIN consists of the C-Score Predictor loss, two contrastive learning losses, and an interaction prediction loss. The final objective of the whole architecture is denoted as:
\begin{equation}
\begin{aligned}
\mathcal{L} = \mathcal{L}_{\text{MSE}}+\mathcal{L}_{\text{CLT}}+\mathcal{L}_{\text{CLD}}+\mathcal{L}_{\text{CE}}
\end{aligned}
\end{equation}

\section{Experiment and Analysis}
\subsection{Experiment Settings}
\subsubsection{Dataset}
\label{dataset}
We train and evaluate our model on two widely used benchmarks ``DUD-E''~\cite{zheng2020predicting} and ``Human''~\cite{liu2015improving}. 

DUD-E dataset consists of $102$ targets across $8$ protein families. On average, each target has $224$ actives (positive drug samples) and over $10,000$ decoys (negative drug samples). The decoys for each target are physically similar but topologically dissimilar to the actives. Overall, there are $22,645$ positive pairs and $1,407,145$ negative pairs in the dataset. The positive samples and randomly chosen equivalent negatives are used to balance the training set for each target. We use a 3-fold cross-validation strategy to evaluate our models following the previous works~\cite{zheng2020predicting,torng2019graph}, the folds are split by targets, where all drugs of the same target belong to the same fold. Besides, targets from the same protein families are kept in the same fold to avoid the impact of homologous proteins.

Using a silico screening process, the negative samples in Human dataset are highly credible. Following the previous works~\cite{chen2020transformercpi,tsubaki2019compound,zheng2020predicting}, we randomly split the datasets as training, validation, and testing sets by the proportion of $80\%,10\%,10\%$, and the ratio of positive and negative samples is $1:1$ in the training set. Finally, Human includes $1,803$ target proteins and $5,423$ interactions.

\subsubsection{Baselines}
We compare our model MIN with the most competitive baselines on DUD-E and Human datasets. We briefly elaborate baselines as follows. 

\textit{\textbf{Docking based}}: (1) Vina~\cite{trott2010autodock} is a program for molecule docking and virtual screening.  

\textit{\textbf{Machine learning based}}: (2) NNscore~\cite{durrant2011nnscore} is a Neural-Network Receptor-Ligand Scoring Function that considers many binding characteristics when predicting affinity; (3) RF-score~\cite{ballester2010machine} is a scoring function that uses Random Forest to implicitly capture binding effects; (4) K nearest neighbors (KNN); (5) Random Forest (RF)~\cite{hall2009weka}; (6) L2-Logistic  (L2)~\cite{fan2008liblinear}.  

\textit{\textbf{Deep learning based}}: (7) GCN~\cite{kipf2016semi} is a scalable approach  based on an efficient variant of convolutional neural networks; (8) 3D-CNN~\cite{ragoza2017protein} takes as input a comprehensive 3D representation of a protein-ligand interaction and automatically learns the key features; (9) PocketGCN~\cite{torng2019graph} is a two-stage graph-convolutional network to learn representations of protein pockets; (10) GraphDTA~\cite{nguyen2019graphdta} predicts drug-target binding affinity using GCN; (11) DrugVQA~\cite{zheng2020predicting} is to predict the interactions by representing proteins with a two-dimensional distance map and drugs with SMILES; (12) GNN~\cite{tsubaki2019compound} combines a GNN for compounds and a CNN for proteins with representation learning; (13) TransformerCPI~\cite{chen2020transformercpi} designs a rigorous label reversal experiment to test whether a model learns true interaction features; (14) GanDTI~\cite{WANG2021107476} employs the compound graph and protein sequence data for both interaction classification and binding affinity prediction tasks; (15) BridgeDPI~\cite{wu2022bridgedpi} introduces a class of nodes named hyper-nodes, which bridge different proteins/drugs to work as protein protein associations and drug-drug associations; (16) AttentionSiteDTI~\cite{yazdani2022attentionsitedti} utilizes protein binding sites along with a self-attention mechanism to address the problem of drug target interaction prediction; (17) DrugCLIP~\cite{gao2024drugclip} reformulates virtual screening as a dense retrieval task and employing contrastive learning to align representations of binding protein pockets and molecules; (18) HyperPCM~\cite{svensson2024hyperpcm} uses HyperNetworks to transfer information between tasks during inference and thus predict drug target interactions on unseen protein targets.

\subsubsection{Metrics}
We evaluate the performance of baselines mainly by the area under the receiver operating characteristic curve (AUC). According to previous studies, we choose corresponding baselines and different evaluation setups for the DUD-E and Human~\cite{zheng2020predicting,tsubaki2019compound,ragoza2017protein}. We report AUC and the ROC enrichment (RE) score at different false positive rate (FPR) thresholds $0.5\%, 1\%, 2\%, 5\%$ for DUD-E dataset. The ROC enrichment is the ratio of the true positive rate (TPR) to the FPR at a given FPR threshold, for example, the upper limit at an FPR of $5\%$ is 20. For Human dataset, we report AUC, the precision score, recall score and F1 score.
\subsubsection{Implementation Details}
Our model is optimized by Adam~\cite{kingma2014adam} with learning rate $0.0001$. The batch size is $256$. The embedding dimension is set to $64$. Transformer in the target branch contains $2$ layers with $8$ attention heads and hidden size $256$, while it is set to $8$ layers with $8$ attention heads and hidden size $512$ in the drug side. GCN in the target branch and drug branch contains $2$ layers and $1$ layer, respectively. We set the dropout for every Transformer and GCN to $0.1$ to avoid overfitting. The conservation threshold is $r_0 = 0.01$. All experiments are carried out on an A40 GPU.

\subsection{Prediction Results}

We compare our proposed model MIN with many DTI prediction approaches on two datasets. The results of MIN are the averages and standard deviations are calculated from the 3-fold cross-validation for DUD-E dataset and from the 3 different random seed runs for Human dataset. $\dag$ means the improvement over the SoTA method is statistically significant ($p < 0.01$). 

The results on DUD-E dataset are shown in Table~\ref{tab:dude}, our model MIN achieves the best performance over all baselines. When comparing MIN with 3D-CNN and AttentionSiteDTI, which both utilize structure information of proteins, MIN achieves an order-of-magnitude improvement. These results indicate that MSA-guided contact map is a good feature extractor to provide structure information and MIN has the ability to make good use of structural information for DTIs prediction. At $RE$ metrics, MIN significantly outperforms the other models, which suggests that MIN is highly effective in identifying true positive interactions at very low false positive rates. The standard deviations for MIN are almost consistently lower than those of other high-performing models like DrugVQA and HyperPCM, indicating that MIN not only achieves better performance but also does so with greater robustness. The statistical significance with a p-value of $0.01$ reinforces that the performance gains observed with MIN are reliable and not due to random variations. Overall, the results clearly demonstrate that MIN achieves state-of-the-art performance across all evaluation metrics on the DUD-E dataset, confirming the efficacy of the multi-channel approach and the use of the C-Score predictor.

\begin{table}[t]
  \centering

    \caption{Performance comparison on the DUD-E dataset. }
    \resizebox{\linewidth}{!}{\begin{tabular}{cccccc}
    \toprule
    \textbf{Models} & \textbf{AUC} & \textbf{0.5\%RE} & \textbf{1\%RE} & \textbf{2\%RE} & \textbf{5\%RE} \\
    \midrule
    \textbf{NNscore} & 0.584 & 4.166 & 2.980  & 2.460  & 1.891 \\
    \textbf{RF-score} & 0.622 & 5.628 & 4.274 & 3.499 & 2.678 \\
    \textbf{Vina} & 0.716 & 9.139 & 7.321 & 5.811 & 4.444 \\
    \textbf{3D-CNN} & 0.868 & 42.559 & 26.655 & 19.363 & 10.710 \\
    \textbf{PocketGCN} & 0.886 & 44.406 & 29.748 & 19.408 & 10.735 \\
    \textbf{DrugVQA} & 0.972 $\pm$ 0.003 & 88.170 $\pm$ 4.88 & 58.710 $\pm$ 2.74 & 35.060 $\pm$ 1.91 & 17.390 $\pm$ 0.94 \\
    \textbf{GanDTI} & 0.943 & 20.960 & 18.134 & 13.617 & 13.018 \\
    \textbf{AttentionSite} & 0.971 & 101.744 & 59.920 & 35.072 & 16.746\\
        \textbf{DrugClip} & 0.966 & 118.10 &67.17 &  37.17 & 16.59 \\
               \textbf{HyperPCM} & 0.982 $\pm$ 0.006 & 183.04 $\pm$ 4.53 & 91.28 $\pm$ 3.35 & 45.62 $\pm$ 2.15 & 17.13 $\pm$ 1.17 \\
    \midrule
    \textbf{Ours: MIN} & \textbf{0.983$^\dag$} & \textbf{197.741$^\dag$}& \textbf{99.563$^\dag$} & \textbf{49.926$^\dag$} & \textbf{19.965$^\dag$} \\
    \midrule
        \textbf{Std of MIN} & 0.002 & 4.731& 2.496 & 1.874 & 0.910 \\
    \bottomrule
    \end{tabular}}%

      \label{tab:dude}%
\end{table}%

\begin{table}[t]
  \centering
    \caption{Performance comparison on the Human dataset.}

   \resizebox{\linewidth}{!}{\begin{tabular}{ccccc}
    \toprule
    \textbf{Models} & \textbf{AUC} & \textbf{Precision} & \textbf{Recall} & \textbf{F1} \\
    \midrule
    \textbf{KNN} & 0.868 & 0.798 & 0.927 & 0.858 \\
    \textbf{RF} & 0.940  & 0.861 & 0.897 & 0.879 \\
    \textbf{L2} & 0.911 & 0.861 & 0.913 & 0.886 \\
    \textbf{GCN} & 0.956 $\pm$ 0.004& 0.862 $\pm$ 0.006& 0.928 $\pm$ 0.010& 0.894 $\pm$ 0.030\\
    \textbf{GNN} & 0.970  & 0.923 & 0.918 & 0.920 \\
    \textbf{GraphDTA} & 0.960 $\pm$ 0.005  & 0.882 $\pm$ 0.040& 0.912 $\pm$ 0.040 & 0.897 $\pm$ 0.052 \\
    \textbf{DrugVQA} & 0.979 $\pm$ 0.005 & 0.954 $\pm$ 0.004 & \textbf{0.961}  $\pm$ 0.003& 0.957 $\pm$ 0.00\\
    \textbf{Transformer CPI} & 0.973 $\pm$ 0.002& 0.916 $\pm$ 0.006 & 0.925 $\pm$ 0.006 & 0.920 $\pm$ 0.004\\
    \textbf{GanDTI} & 0.981 & 0.958 & 0.919 & 0.938 \\
    \textbf{BridgeDPI} & 0.979 & 0.929 & 0.936 & 0.93 \\
     \textbf{DrugClip} & 0.977 & 0.961 & 0.958 & 0.961 \\
      \textbf{HyperPCM} & 0.976 & 0.918 & 0.928 & 0.921\\
    \midrule
    \textbf{Ours: MIN} & \textbf{0.984$^\dag$}& \textbf{0.983$^\dag$} & 0.952 & \textbf{0.965$^\dag$} \\
       \midrule
    \textbf{Std of MIN} & 0.0031& 0.0027 & 0.0029 & 0.0028 \\
    \bottomrule
    \end{tabular}%
    }
      \label{tab:human}%
\end{table}%

We compare MIN with $12$ DTI prediction approaches on the Human dataset. As Table~\ref{tab:human} shows, MIN outperforms all baselines in AUC and far exceeds any other baselines with \text{Recall} score and \text{F1} score with relatively low standard deviations. Although DrugVQA has the highest recall score of $0.961$, MIN’s recall score is close at $0.952$, still placing it among the top performers. This slight trade-off in recall is offset by the overall gain in F1 score and AUC. MIN achieves an F1 score of $0.965$, surpassing all other models, including DrugVQA with an F1 score of $0.957$ and DrugClip with $0.961$. This demonstrates that MIN provides the best balance between precision and recall, ensuring reliable predictions. Overall, experiments on the Human dataset demonstrate that MIN consistently outperforms all baselines across multiple evaluation metrics.

\subsection{Ablation Study}
To understand the contributions of different components in MIN, we conduct a number of ablation experiments where we remove C-score Predictor (w/o CS), contrastive learning (w/o CL),  structure-agnostic channel (w/o SAgC),  structure-aware channel (w/o SAwC),  extended-mixture channel (w/o EMC), multi-channel interaction module (w/o MCI) and their combinations from the original MIN architecture, respectively. We also remove MSA-guided contact graph from protein inputs (w/o PG) and graph from drug inputs (w/o DG) respectively to explore the value of structure information. 

\begin{table}[htbp]
  \centering
    \caption{Ablation study of MIN architecture on DUD-E.}
    \begin{tabular}{lc}
    \toprule
    \multicolumn{1}{c}{\textbf{Models}} & \textbf{AUC} \\
        \midrule
    \textbf{w/o} SAgC & 0.978  \\
        \midrule
    \textbf{w/o} SAwC & 0.977  \\
    \midrule
    \textbf{w/o} EMC & 0.974  \\
    \midrule
    \textbf{w/o} MCI & 0.973  \\
    \midrule
    \textbf{w/o} CL & 0.970  \\
    \midrule
    \textbf{w/o} CS & 0.968  \\
    \midrule
    \multicolumn{1}{p{17em}}{\textbf{w/o} MCI \textbf{w/o} PG} & 0.963  \\
    \midrule
    \multicolumn{1}{p{17em}}{\textbf{w/o} MCI \textbf{w/o} PG \textbf{w/o} CL} & 0.959  \\
    \midrule
    \multicolumn{1}{p{17em}}{\textbf{w/o} MCI \textbf{w/o} PG \textbf{w/o} DG} & 0.955  \\
    \midrule
    \multicolumn{1}{p{17em}}{\textbf{w/o} MCI \textbf{w/o} PG \textbf{w/o} DG \textbf{w/o} CS} & 0.951  \\
    \midrule
    \textbf{Full model} & \textbf{0.983 } \\
    \bottomrule
    \end{tabular}%

    \label{tab:ablation}
\end{table}%

The results are shown in Table~\ref{tab:ablation}. We have the following observations. We could see that every channel in the multi-channel interaction module is valuable. MIN without contrastive learning performs worse: AUC score is $0.970$, which is lower than the full model. This is due to contrastive learning could help to unify two representations. MIN without the C-Score Predictor means the original proteins are not distilled. The AUC of MIN without the C-Score Predictor is down to $0.968$, which indicates the most prominent effectiveness of the C-Score Predictor as a single factor. When we use an MLP to replace MCI, the AUC is $0.01$ lower than that in the full MIN. It is clear that the multi-channel interaction mechanism can enhance the performance of DTI predictions. We also note that when the combinations are ablated from the full model, the performance drops severely. In the last ablation experiment, we remove all contribution points in MIN, where the AUC value is lower than all other ablation variants.

\subsection{Case Study}

We conduct a case study to answer the following two questions about MIN: 

\vspace{0.6em}
\noindent\textbf{Q1: Does C-Score Predictor has ability to distill long proteins effectively?}

Here we choose a long protein case to provide the prediction result of C-Score Predictor. For protein Mitogen-activated protein kinase kinase kinase 5 (UniProt ID:Q99683) and compound Adenosine (PubChem CID:60961), they could interact with each other. Without C-Score Predictor, the predicted possibility of $\hat{y}=1$ is 0.5, which means that the model can't give a certain prediction of whether interaction exists. C-Score Predictor selects $588$ residues to preserve while the original sequence has $1,608$ residues. As expected, C-Score Predictor leads to a better result with a pretty sure possibility $0.999$. 

Besides, in order to present the general behavior of C-Score Predictor, we compare the prediction results changed along with protein length using following strategies. 
\begin{itemize}
    \item \textbf{w/o distill}: Removing C-Score Predictor from MIN, which means that original protein sequences are encoded without distillation; 

\item \textbf{Distill by C-Score}: Leveraging conservation score to select conservative residues directly under the given threshold $r_0$; 

\item \textbf{Distill by C-Score Predictor}: Training a conservative residue recognizer, which is also influenced by DTI prediction objective. 
\end{itemize}

\begin{figure}[ht]
  \centering
\includegraphics[width=8cm]{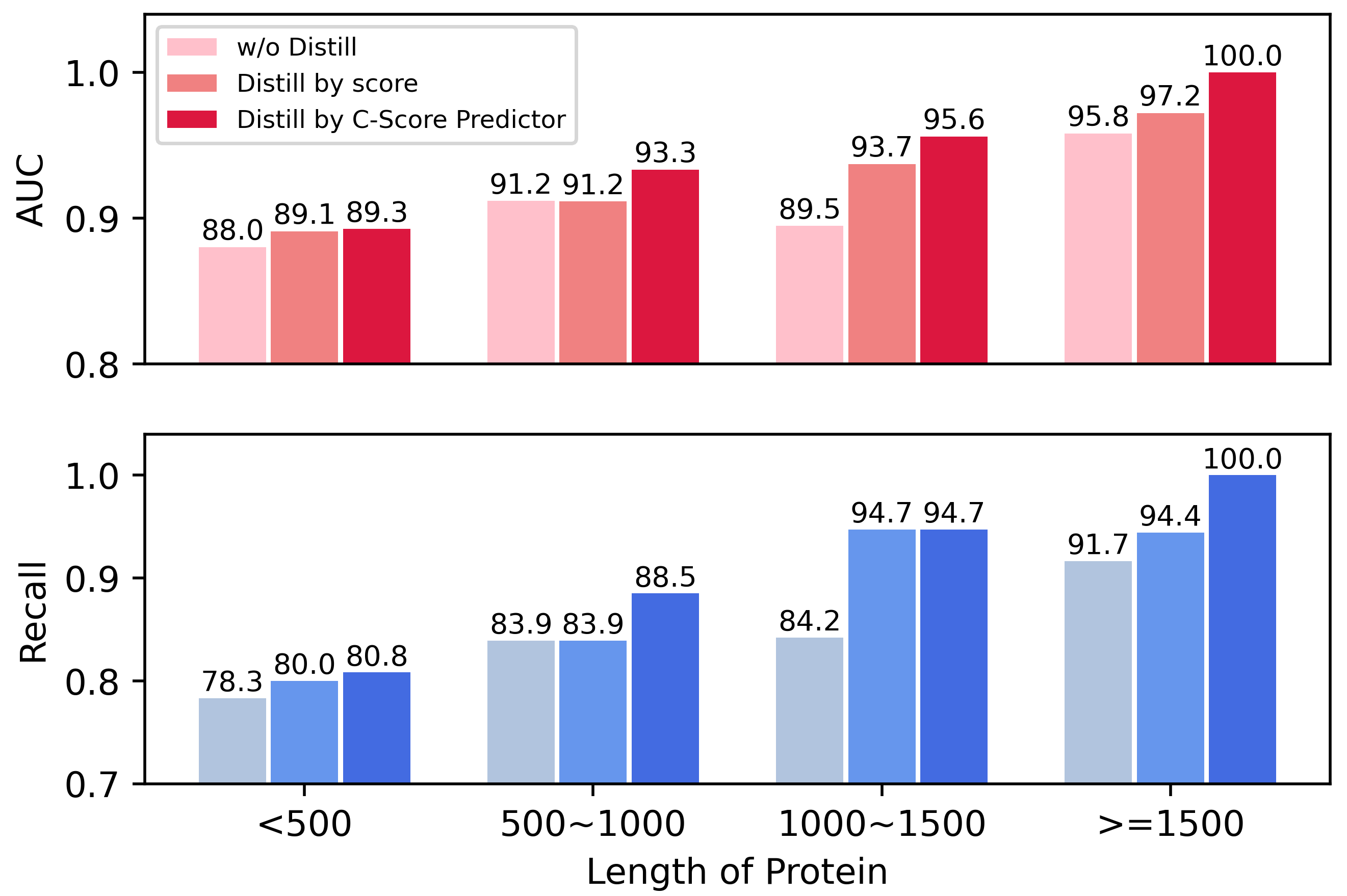}
  \caption{Performance comparison between C-Score Predictor with other methods changed along with protein length on DUD-E, which indicates C-Score Predictor has ability to distill long proteins effectively.}
  \label{fig:class4}
\end{figure}

The comparison results are shown in Figure~\ref{fig:class4}.  We can see that the performance of w/o Distill is the worst, which verify that conservation score is effective information for the DTI task. Besides, it is easy to find that with the increase of protein length C-Score Predictor performs better than Distill by C-Score directly. It may mistakenly remove some residues that are conservative but crucial for DTI prediction, as conservation is just an essential factor but not the determinant. So C-Score Predictor has the ability to select vital residues based on both DTI prediction objective and biological meaningful conservation.

\vspace{0.6em}
\noindent\textbf{Q2: Could MIN learn binding pattern at a detailed level?}

\begin{figure}[ht]
  \centering
\includegraphics[width=6.5cm]{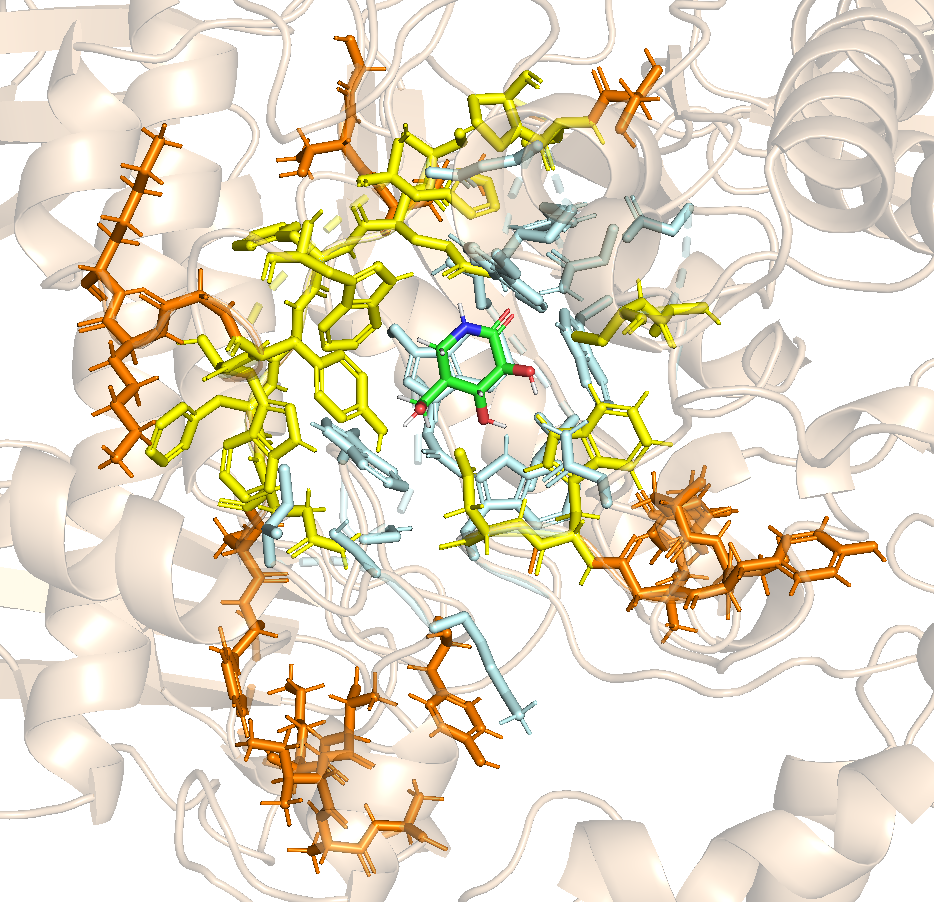}
  \caption{The comparison of the real binding pocket and prediction binding pocket. Among 37 residues in the binding pocket, 16 overlap residues between real and prediction are painted yellow.}
  \label{fig:whole}
\end{figure}

\begin{figure}[ht]
  \centering
\includegraphics[width=6.5cm]{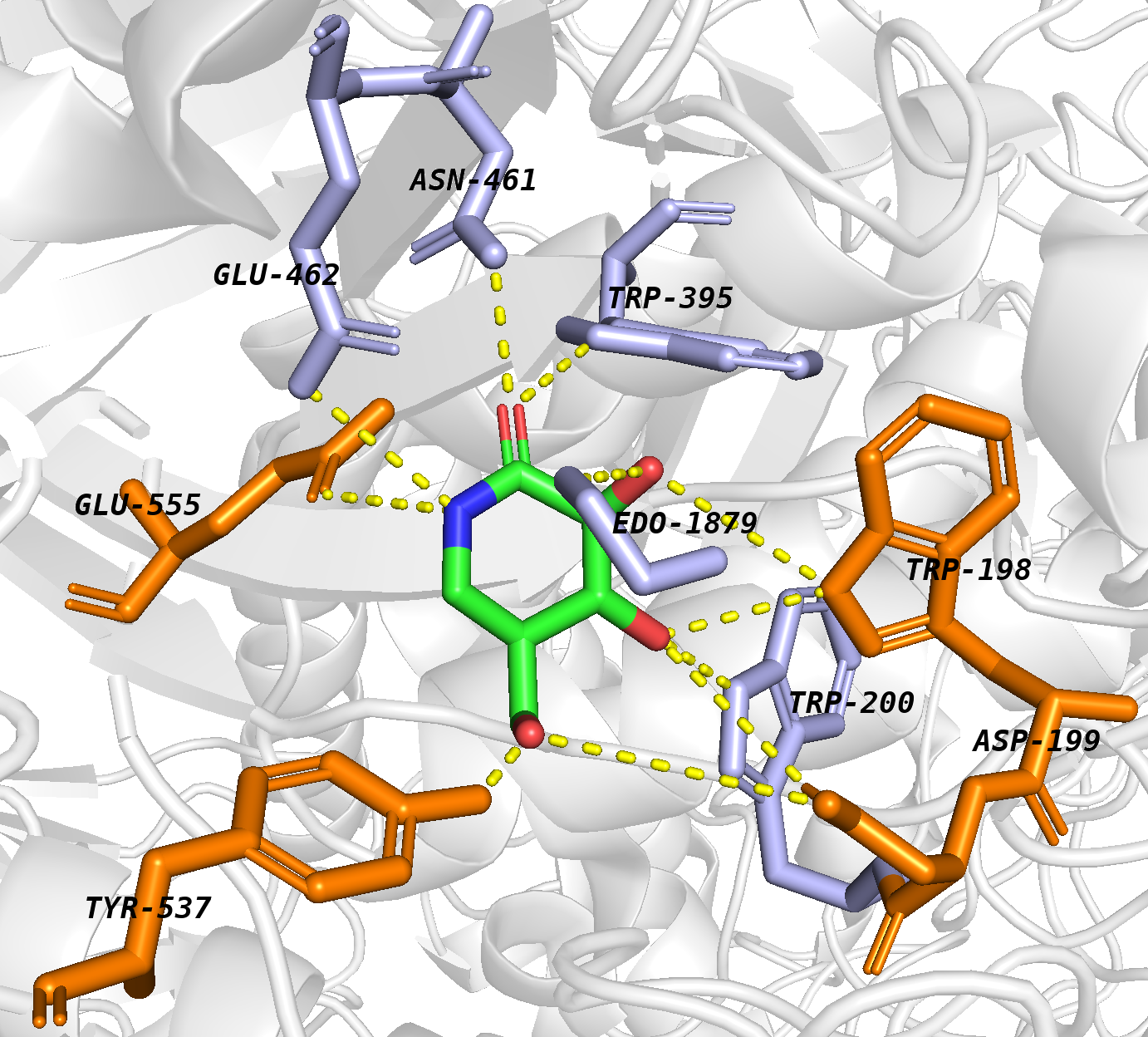}
  \caption{The comparison of real binding residues and prediction binding residues. Among the $9$ real binding residues, $4$ of them in orange are in the prediction list.}
  \label{fig:detail}
\end{figure}

To explore if MIN could learn the binding pattern in detail, a case PDBID: $2vjx$ from PDBBind~\cite{wang2005pdbbind} is selected to present. We use PyMOL software to draw the following pictures~\footnote{https://pymol.org/2/}. There are $37$ residues in the real binding pocket and among them $9$ residues are real binding positions. MIN trained on DUD-E is used to test this case. We calculate the difference of logit values $f([c1, c2, c3])$ by Equation~\ref{logit} between masking a residue or not to explore the importance of this residue. This process is performed for every residue, then we rank the logit differences and take out the top $37$ residues (same as the pocket length) as prediction binding pocket by MIN. Figure~\ref{fig:whole} shows the comparison of real binding pocket and prediction binding pocket, in which the ligand is painted green, the overlap residues between real and prediction are printed yellow, the real pocket residues are presented in lightblue while the prediction pocket residues are painted orange. Among the $37$ selected residues, $16$ residues locate in the real binding pocket, which indicates that MIN has the ability to find binding pocket with high accuracy. Besides, Figure~\ref{fig:detail} presents the details of binding residues. There are $9$ binding residues in this case, among them $4$ residues painted orange (TRP-198, ASP-199, TYR-537, GLU-555) are selected by MIN. This supports that our model could not only locate the binding pocket but also find the binding residues precisely. These observations suggest that our model has a good generalization ability to predict the case from other datasets and to make prediction at a very detailed level.

\section{Conclusion}
Our primary goal in this paper is to predict drug-target interactions, and we introduce a novel model called MIN. MIN leverages sequence representations and structure representations for both target and drug. We design a C-Score Predictor to improve the prediction efficiency and performance by distilling the long protein sequences. Besides, a multi-channel interaction module enhanced by contrastive learning is proposed. This module extracts interaction patterns from three levels to make predictions. Experimental evaluations show that our model achieves good performance on two datasets, and it may open a new avenue for drug discovery.

\section{Acknowledgement}
Shuqi Li is supported by the Fundamental Research Funds for the Central Universities, and the Research Funds of Renmin University of China (NO.297522615221).

\bibliographystyle{IEEEtranN}
\bibliography{ref.bib}

% Generated by IEEEtranN.bst, version: 1.14 (2015/08/26)
\begin{thebibliography}{55}
\providecommand{\natexlab}[1]{#1}
\providecommand{\url}[1]{#1}
\csname url@samestyle\endcsname
\providecommand{\newblock}{\relax}
\providecommand{\bibinfo}[2]{#2}
\providecommand{\BIBentrySTDinterwordspacing}{\spaceskip=0pt\relax}
\providecommand{\BIBentryALTinterwordstretchfactor}{4}
\providecommand{\BIBentryALTinterwordspacing}{\spaceskip=\fontdimen2\font plus
\BIBentryALTinterwordstretchfactor\fontdimen3\font minus \fontdimen4\font\relax}
\providecommand{\BIBforeignlanguage}[2]{{%
\expandafter\ifx\csname l@#1\endcsname\relax
\typeout{** WARNING: IEEEtranN.bst: No hyphenation pattern has been}%
\typeout{** loaded for the language `#1'. Using the pattern for}%
\typeout{** the default language instead.}%
\else
\language=\csname l@#1\endcsname
\fi
#2}}
\providecommand{\BIBdecl}{\relax}
\BIBdecl

\bibitem[Haggarty et~al.(2003)Haggarty, Koeller, Wong, Butcher, and Schreiber]{haggarty2003multidimensional}
S.~J. Haggarty, K.~M. Koeller, J.~C. Wong, R.~A. Butcher, and S.~L. Schreiber, ``Multidimensional chemical genetic analysis of diversity-oriented synthesis-derived deacetylase inhibitors using cell-based assays,'' \emph{Chemistry \& biology}, vol.~10, no.~5, pp. 383--396, 2003.

\bibitem[Chen et~al.(2016)Chen, Yan, Zhang, Zhang, Dai, Yin, and Zhang]{chen2016drug}
X.~Chen, C.~C. Yan, X.~Zhang, X.~Zhang, F.~Dai, J.~Yin, and Y.~Zhang, ``Drug--target interaction prediction: databases, web servers and computational models,'' \emph{Briefings in bioinformatics}, vol.~17, no.~4, pp. 696--712, 2016.

\bibitem[Keiser et~al.(2007)Keiser, Roth, Armbruster, Ernsberger, Irwin, and Shoichet]{keiser2007relating}
M.~J. Keiser, B.~L. Roth, B.~N. Armbruster, P.~Ernsberger, J.~J. Irwin, and B.~K. Shoichet, ``Relating protein pharmacology by ligand chemistry,'' \emph{Nature biotechnology}, vol.~25, no.~2, pp. 197--206, 2007.

\bibitem[Keiser et~al.(2009)Keiser, Setola, Irwin, Laggner, Abbas, Hufeisen, Jensen, Kuijer, Matos, Tran, et~al.]{keiser2009predicting}
M.~J. Keiser, V.~Setola, J.~J. Irwin, C.~Laggner, A.~I. Abbas, S.~J. Hufeisen, N.~H. Jensen, M.~B. Kuijer, R.~C. Matos, T.~B. Tran \emph{et~al.}, ``Predicting new molecular targets for known drugs,'' \emph{Nature}, vol. 462, no. 7270, pp. 175--181, 2009.

\bibitem[Cheng et~al.(2007)Cheng, Coleman, Smyth, Cao, Soulard, Caffrey, Salzberg, and Huang]{cheng2007structure}
A.~C. Cheng, R.~G. Coleman, K.~T. Smyth, Q.~Cao, P.~Soulard, D.~R. Caffrey, A.~C. Salzberg, and E.~S. Huang, ``Structure-based maximal affinity model predicts small-molecule druggability,'' \emph{Nature biotechnology}, vol.~25, no.~1, pp. 71--75, 2007.

\bibitem[Donald(2011)]{donald2011algorithms}
B.~R. Donald, \emph{Algorithms in structural molecular biology}.\hskip 1em plus 0.5em minus 0.4em\relax MIT Press, 2011.

\bibitem[Fu and Meiler(2018)]{fu2018predictive}
D.~Y. Fu and J.~Meiler, ``Predictive power of different types of experimental restraints in small molecule docking: A review,'' \emph{Journal of chemical information and modeling}, vol.~58, no.~2, pp. 225--233, 2018.

\bibitem[Alaimo et~al.(2016)Alaimo, Giugno, and Pulvirenti]{alaimo2016recommendation}
S.~Alaimo, R.~Giugno, and A.~Pulvirenti, ``Recommendation techniques for drug--target interaction prediction and drug repositioning,'' in \emph{Data Mining Techniques for the Life Sciences}.\hskip 1em plus 0.5em minus 0.4em\relax Springer, 2016, pp. 441--462.

\bibitem[Wen et~al.(2017)Wen, Zhang, Niu, Sha, Yang, Yun, and Lu]{wen2017deep}
M.~Wen, Z.~Zhang, S.~Niu, H.~Sha, R.~Yang, Y.~Yun, and H.~Lu, ``Deep-learning-based drug--target interaction prediction,'' \emph{Journal of proteome research}, vol.~16, no.~4, pp. 1401--1409, 2017.

\bibitem[Bleakley and Yamanishi(2009)]{Bleakley2009Supervised}
K.~Bleakley and Y.~Yamanishi, ``Supervised prediction of drug--target interactions using bipartite local models,'' \emph{Bioinformatics}, vol.~25, no.~18, pp. 2397--2403, 2009.

\bibitem[Weininger(1988)]{weininger1988smiles}
D.~Weininger, ``Smiles, a chemical language and information system. 1. introduction to methodology and encoding rules,'' \emph{Journal of chemical information and computer sciences}, vol.~28, no.~1, pp. 31--36, 1988.

\bibitem[Ragoza et~al.(2017)Ragoza, Hochuli, Idrobo, Sunseri, and Koes]{ragoza2017protein}
M.~Ragoza, J.~Hochuli, E.~Idrobo, J.~Sunseri, and D.~R. Koes, ``Protein--ligand scoring with convolutional neural networks,'' \emph{Journal of chemical information and modeling}, vol.~57, no.~4, pp. 942--957, 2017.

\bibitem[Stepniewska-Dziubinska et~al.(2018)Stepniewska-Dziubinska, Zielenkiewicz, and Siedlecki]{stepniewska2018development}
M.~M. Stepniewska-Dziubinska, P.~Zielenkiewicz, and P.~Siedlecki, ``Development and evaluation of a deep learning model for protein--ligand binding affinity prediction,'' \emph{Bioinformatics}, vol.~34, no.~21, pp. 3666--3674, 2018.

\bibitem[Torng and Altman(2019)]{torng2019graph}
W.~Torng and R.~B. Altman, ``Graph convolutional neural networks for predicting drug-target interactions,'' \emph{Journal of chemical information and modeling}, vol.~59, no.~10, pp. 4131--4149, 2019.

\bibitem[Zheng et~al.(2020)Zheng, Li, Chen, Xu, and Yang]{zheng2020predicting}
S.~Zheng, Y.~Li, S.~Chen, J.~Xu, and Y.~Yang, ``Predicting drug--protein interaction using quasi-visual question answering system,'' \emph{Nature Machine Intelligence}, vol.~2, no.~2, pp. 134--140, 2020.

\bibitem[{\"O}zt{\"u}rk et~al.(2018){\"O}zt{\"u}rk, {\"O}zg{\"u}r, and Ozkirimli]{ozturk2018deepdta}
H.~{\"O}zt{\"u}rk, A.~{\"O}zg{\"u}r, and E.~Ozkirimli, ``Deepdta: deep drug--target binding affinity prediction,'' \emph{Bioinformatics}, vol.~34, no.~17, pp. i821--i829, 2018.

\bibitem[Karimi et~al.(2019)Karimi, Wu, Wang, and Shen]{karimi2019deepaffinity}
M.~Karimi, D.~Wu, Z.~Wang, and Y.~Shen, ``Deepaffinity: interpretable deep learning of compound--protein affinity through unified recurrent and convolutional neural networks,'' \emph{Bioinformatics}, vol.~35, no.~18, pp. 3329--3338, 2019.

\bibitem[Huang et~al.(2020)Huang, Fu, Glass, Zitnik, Xiao, and Sun]{2020DeepPurpose}
K.~Huang, T.~Fu, L.~Glass, M.~Zitnik, C.~Xiao, and J.~Sun, ``Deeppurpose: a deep learning library for drug-target interaction prediction,'' \emph{Bioinformatics}, 2020.

\bibitem[Li et~al.(2020)Li, Wan, Shu, Jiang, Zhao, and Zeng]{li2020monn}
S.~Li, F.~Wan, H.~Shu, T.~Jiang, D.~Zhao, and J.~Zeng, ``Monn: a multi-objective neural network for predicting compound-protein interactions and affinities,'' \emph{Cell Systems}, vol.~10, no.~4, pp. 308--322, 2020.

\bibitem[Kipf and Welling(2016)]{kipf2016semi}
T.~N. Kipf and M.~Welling, ``Semi-supervised classification with graph convolutional networks,'' \emph{arXiv preprint arXiv:1609.02907}, 2016.

\bibitem[Durrant and Mccammon(2011)]{2011NNScore}
J.~D. Durrant and J.~A. Mccammon, ``Nnscore 2.0: A neural-network receptor–ligand scoring function,'' \emph{Journal of Chemical Information \& Modeling}, vol.~51, no.~11, p. 2897, 2011.

\bibitem[Cai et~al.(2021)Cai, Lim, Abbu, Qiu, Nussinov, and Xie]{cai2021msa}
T.~Cai, H.~Lim, K.~A. Abbu, Y.~Qiu, R.~Nussinov, and L.~Xie, ``Msa-regularized protein sequence transformer toward predicting genome-wide chemical-protein interactions: Application to gpcrome deorphanization,'' \emph{Journal of chemical information and modeling}, vol.~61, no.~4, pp. 1570--1582, 2021.

\bibitem[Chen et~al.(2020)Chen, Tan, Wang, Zhong, Liu, Yang, Luo, Chen, Jiang, and Zheng]{chen2020transformercpi}
L.~Chen, X.~Tan, D.~Wang, F.~Zhong, X.~Liu, T.~Yang, X.~Luo, K.~Chen, H.~Jiang, and M.~Zheng, ``Transformercpi: improving compound--protein interaction prediction by sequence-based deep learning with self-attention mechanism and label reversal experiments,'' \emph{Bioinformatics}, vol.~36, no.~16, pp. 4406--4414, 2020.

\bibitem[Wang et~al.(2021)Wang, Shan, Zhao, and Zuo]{WANG2021107476}
\BIBentryALTinterwordspacing
S.~Wang, P.~Shan, Y.~Zhao, and L.~Zuo, ``Gandti: A multi-task neural network for drug-target interaction prediction,'' \emph{Computational Biology and Chemistry}, vol.~92, p. 107476, 2021. [Online]. Available: \url{https://www.sciencedirect.com/science/article/pii/S1476927121000438}
\BIBentrySTDinterwordspacing

\bibitem[Wu et~al.(2022)Wu, Gao, Zeng, Zhang, and Li]{wu2022bridgedpi}
Y.~Wu, M.~Gao, M.~Zeng, J.~Zhang, and M.~Li, ``Bridgedpi: a novel graph neural network for predicting drug--protein interactions,'' \emph{Bioinformatics}, vol.~38, no.~9, pp. 2571--2578, 2022.

\bibitem[Yazdani-Jahromi et~al.(2022)Yazdani-Jahromi, Yousefi, Tayebi, Kolanthai, Neal, Seal, and Garibay]{yazdani2022attentionsitedti}
M.~Yazdani-Jahromi, N.~Yousefi, A.~Tayebi, E.~Kolanthai, C.~J. Neal, S.~Seal, and O.~O. Garibay, ``Attentionsitedti: an interpretable graph-based model for drug-target interaction prediction using nlp sentence-level relation classification,'' \emph{Briefings in Bioinformatics}, 2022.

\bibitem[Moon et~al.(2022)Moon, Zhung, Yang, Lim, and Kim]{moon2022pignet}
S.~Moon, W.~Zhung, S.~Yang, J.~Lim, and W.~Y. Kim, ``Pignet: a physics-informed deep learning model toward generalized drug--target interaction predictions,'' \emph{Chemical Science}, vol.~13, no.~13, pp. 3661--3673, 2022.

\bibitem[Song et~al.(2022)Song, Zhang, Ding, Rodriguez-Paton, Wang, and Wang]{SONG2022269}
\BIBentryALTinterwordspacing
T.~Song, X.~Zhang, M.~Ding, A.~Rodriguez-Paton, S.~Wang, and G.~Wang, ``Deepfusion: A deep learning based multi-scale feature fusion method for predicting drug-target interactions,'' \emph{Methods}, vol. 204, pp. 269--277, 2022. [Online]. Available: \url{https://www.sciencedirect.com/science/article/pii/S1046202322000378}
\BIBentrySTDinterwordspacing

\bibitem[Dehghan et~al.(2023)Dehghan, Razzaghi, Abbasi, and Gharaghani]{dehghan2023tripletmultidti}
A.~Dehghan, P.~Razzaghi, K.~Abbasi, and S.~Gharaghani, ``Tripletmultidti: multimodal representation learning in drug-target interaction prediction with triplet loss function,'' \emph{Expert Systems with Applications}, vol. 232, p. 120754, 2023.

\bibitem[Feng et~al.(2024)Feng, Zhang, Deng, and Xiong]{feng2024gcardti}
Y.~Feng, Y.~Zhang, Z.~Deng, and M.~Xiong, ``Gcardti: Drug--target interaction prediction based on a hybrid mechanism in drug selfies,'' \emph{Quantitative Biology}, 2024.

\bibitem[Liu et~al.(2024)Liu, Wu, Wang, Deng, and Zhou]{liu2024higraphdti}
B.~Liu, S.~Wu, J.~Wang, X.~Deng, and A.~Zhou, ``Higraphdti: Hierarchical graph representation learning for drug-target interaction prediction,'' in \emph{Joint European Conference on Machine Learning and Knowledge Discovery in Databases}.\hskip 1em plus 0.5em minus 0.4em\relax Springer, 2024, pp. 354--370.

\bibitem[Zhao et~al.(2024)Zhao, Wang, and Shi]{zhao2024pocketdta}
L.~Zhao, H.~Wang, and S.~Shi, ``Pocketdta: an advanced multimodal architecture for enhanced prediction of drug- target affinity from 3d structural data of target binding pockets,'' \emph{Bioinformatics}, vol.~40, no.~10, p. btae594, 2024.

\bibitem[Gao et~al.(2024)Gao, Qiang, Tan, Jia, Ren, Lu, Liu, Ma, and Lan]{gao2024drugclip}
B.~Gao, B.~Qiang, H.~Tan, Y.~Jia, M.~Ren, M.~Lu, J.~Liu, W.-Y. Ma, and Y.~Lan, ``Drugclip: Contrasive protein-molecule representation learning for virtual screening,'' \emph{Advances in Neural Information Processing Systems}, vol.~36, 2024.

\bibitem[Svensson et~al.(2024)Svensson, Hoedt, Hochreiter, and Klambauer]{svensson2024hyperpcm}
E.~Svensson, P.-J. Hoedt, S.~Hochreiter, and G.~Klambauer, ``Hyperpcm: Robust task-conditioned modeling of drug--target interactions,'' \emph{Journal of Chemical Information and Modeling}, vol.~64, no.~7, pp. 2539--2553, 2024.

\bibitem[Mirdita et~al.(2016)Mirdita, von den Driesch, Galiez, Martin, Söding, and Steinegger]{10.1093/nar/gkw1081}
\BIBentryALTinterwordspacing
M.~Mirdita, L.~von den Driesch, C.~Galiez, M.~J. Martin, J.~Söding, and M.~Steinegger, ``{Uniclust databases of clustered and deeply annotated protein sequences and alignments},'' \emph{Nucleic Acids Research}, vol.~45, no.~D1, pp. D170--D176, 11 2016. [Online]. Available: \url{https://doi.org/10.1093/nar/gkw1081}
\BIBentrySTDinterwordspacing

\bibitem[Capra and Singh(2007)]{msa}
J.~A. Capra and M.~Singh, ``{Predicting functionally important residues from sequence conservation},'' \emph{Bioinformatics}, vol.~23, no.~15, pp. 1875--1882, 05 2007.

\bibitem[Vaswani et~al.(2017)Vaswani, Shazeer, Parmar, Uszkoreit, Jones, Gomez, Kaiser, and Polosukhin]{vaswani2017attention}
A.~Vaswani, N.~Shazeer, N.~Parmar, J.~Uszkoreit, L.~Jones, A.~N. Gomez, {\L}.~Kaiser, and I.~Polosukhin, ``Attention is all you need,'' in \emph{Advances in neural information processing systems}, 2017, pp. 5998--6008.

\bibitem[Guharoy and Chakrabarti(2005)]{guharoy2005conservation}
M.~Guharoy and P.~Chakrabarti, ``Conservation and relative importance of residues across protein-protein interfaces,'' \emph{Proceedings of the National Academy of Sciences}, vol. 102, no.~43, pp. 15\,447--15\,452, 2005.

\bibitem[Wang et~al.(2004)Wang, Fang, Lu, and Wang]{wang2004pdbbind}
R.~Wang, X.~Fang, Y.~Lu, and S.~Wang, ``The pdbbind database: Collection of binding affinities for protein- ligand complexes with known three-dimensional structures,'' \emph{Journal of medicinal chemistry}, vol.~47, no.~12, pp. 2977--2980, 2004.

\bibitem[Ben~Chorin et~al.(2020)Ben~Chorin, Masrati, Kessel, Narunsky, Sprinzak, Lahav, Ashkenazy, and Ben-Tal]{ben2020consurf}
A.~Ben~Chorin, G.~Masrati, A.~Kessel, A.~Narunsky, J.~Sprinzak, S.~Lahav, H.~Ashkenazy, and N.~Ben-Tal, ``Consurf-db: An accessible repository for the evolutionary conservation patterns of the majority of pdb proteins,'' \emph{Protein Science}, vol.~29, no.~1, pp. 258--267, 2020.

\bibitem[Goldenberg et~al.(2009)Goldenberg, Erez, Nimrod, and Ben-Tal]{goldenberg2009consurf}
O.~Goldenberg, E.~Erez, G.~Nimrod, and N.~Ben-Tal, ``The consurf-db: pre-calculated evolutionary conservation profiles of protein structures,'' \emph{Nucleic acids research}, vol.~37, no. suppl\_1, pp. D323--D327, 2009.

\bibitem[Zhu et~al.(2021)Zhu, Xia, Qin, Zhou, Li, and Liu]{zhu2021dual}
J.~Zhu, Y.~Xia, T.~Qin, W.~Zhou, H.~Li, and T.-Y. Liu, ``Dual-view molecule pre-training,'' \emph{arXiv preprint arXiv:2106.10234}, 2021.

\bibitem[Jiang et~al.(2023)Jiang, Huang, and Huang]{jiang2023adaptive}
Y.~Jiang, C.~Huang, and L.~Huang, ``Adaptive graph contrastive learning for recommendation,'' in \emph{Proceedings of the 29th ACM SIGKDD conference on knowledge discovery and data mining}, 2023, pp. 4252--4261.

\bibitem[Xiao et~al.(2024)Xiao, Zhu, Chen, and Wang]{xiao2024simple}
T.~Xiao, H.~Zhu, Z.~Chen, and S.~Wang, ``Simple and asymmetric graph contrastive learning without augmentations,'' \emph{Advances in Neural Information Processing Systems}, vol.~36, 2024.

\bibitem[He et~al.(2020)He, Fan, Wu, Xie, and Girshick]{he2020momentum}
K.~He, H.~Fan, Y.~Wu, S.~Xie, and R.~Girshick, ``Momentum contrast for unsupervised visual representation learning,'' in \emph{Proceedings of the IEEE/CVF Conference on Computer Vision and Pattern Recognition}, 2020, pp. 9729--9738.

\bibitem[Liu et~al.(2015)Liu, Sun, Guan, Zheng, and Zhou]{liu2015improving}
H.~Liu, J.~Sun, J.~Guan, J.~Zheng, and S.~Zhou, ``Improving compound--protein interaction prediction by building up highly credible negative samples,'' \emph{Bioinformatics}, vol.~31, no.~12, pp. i221--i229, 2015.

\bibitem[Tsubaki et~al.(2019)Tsubaki, Tomii, and Sese]{tsubaki2019compound}
M.~Tsubaki, K.~Tomii, and J.~Sese, ``Compound--protein interaction prediction with end-to-end learning of neural networks for graphs and sequences,'' \emph{Bioinformatics}, vol.~35, no.~2, pp. 309--318, 2019.

\bibitem[Trott and Olson(2010)]{trott2010autodock}
O.~Trott and A.~J. Olson, ``Autodock vina: improving the speed and accuracy of docking with a new scoring function, efficient optimization, and multithreading,'' \emph{Journal of computational chemistry}, vol.~31, no.~2, pp. 455--461, 2010.

\bibitem[Durrant and McCammon(2011)]{durrant2011nnscore}
J.~D. Durrant and J.~A. McCammon, ``Nnscore 2.0: a neural-network receptor--ligand scoring function,'' \emph{Journal of chemical information and modeling}, vol.~51, no.~11, pp. 2897--2903, 2011.

\bibitem[Ballester and Mitchell(2010)]{ballester2010machine}
P.~J. Ballester and J.~B. Mitchell, ``A machine learning approach to predicting protein--ligand binding affinity with applications to molecular docking,'' \emph{Bioinformatics}, vol.~26, no.~9, pp. 1169--1175, 2010.

\bibitem[Hall et~al.(2009)Hall, Frank, Holmes, Pfahringer, Reutemann, and Witten]{hall2009weka}
M.~Hall, E.~Frank, G.~Holmes, B.~Pfahringer, P.~Reutemann, and I.~H. Witten, ``The weka data mining software: an update,'' \emph{ACM SIGKDD explorations newsletter}, vol.~11, no.~1, pp. 10--18, 2009.

\bibitem[Fan et~al.(2008)Fan, Chang, Hsieh, Wang, and Lin]{fan2008liblinear}
R.-E. Fan, K.-W. Chang, C.-J. Hsieh, X.-R. Wang, and C.-J. Lin, ``Liblinear: A library for large linear classification,'' \emph{the Journal of machine Learning research}, vol.~9, pp. 1871--1874, 2008.

\bibitem[Nguyen et~al.(2019)Nguyen, Le, and Venkatesh]{nguyen2019graphdta}
T.~Nguyen, H.~Le, and S.~Venkatesh, ``Graphdta: prediction of drug--target binding affinity using graph convolutional networks,'' \emph{BioRxiv}, p. 684662, 2019.

\bibitem[Kingma and Ba(2014)]{kingma2014adam}
D.~P. Kingma and J.~Ba, ``Adam: A method for stochastic optimization,'' \emph{arXiv preprint arXiv:1412.6980}, 2014.

\bibitem[Wang et~al.(2005)Wang, Fang, Lu, Yang, and Wang]{wang2005pdbbind}
R.~Wang, X.~Fang, Y.~Lu, C.-Y. Yang, and S.~Wang, ``The pdbbind database: methodologies and updates,'' \emph{Journal of medicinal chemistry}, vol.~48, no.~12, pp. 4111--4119, 2005.

\end{thebibliography}

\begin{IEEEbiographynophoto}
{Shuqi Li} is currently working toward a Ph.D. degree with the Gaoling School of Artificial Intelligence (GSAI), Renmin University of China (RUC). Her research mainly focuses on Natural Language Processing, Text Mining, AI for Science and FinTech. She has published papers with conferences such as AAAI, ACL and NeurIPS.
\end{IEEEbiographynophoto}

\begin{IEEEbiographynophoto}
{Shufang Xie} received his BE degree in computer science and technology from the Southeast University in 2016. He is currently working towards the Ph.D. degree with the Renmin University of China. His research main focus on machine learning and AI for science.
\end{IEEEbiographynophoto}

\begin{IEEEbiographynophoto}
{Hongda Sun} is a Ph.D. candidate of Gaoling School of Artificial Intelligence (GSAI), Renmin University of China. In 2020, he was recommended for GSAI as a direct doctoral student, supervised by Dr. Rui Yan. His research works have been published in NeurIPS and AAAI. Now his major interests are in AI for science, machine learning and natural language processing.
\end{IEEEbiographynophoto}

\begin{IEEEbiographynophoto}
{Yuhan Chen} is currently working toward the M.S. degree with the Renmin University of China (RUC). Her research mainly focus on Natural Language Processing, Dialogue System, AI for Science and FinTech. She has published papers with conferences such as AAAI and ACL.
\end{IEEEbiographynophoto}

\begin{IEEEbiographynophoto}
{Tao Qin} is a Senior Principal Researcher/Manager  at Microsoft Research AI4Science. His research interests include deep learning (with applications to machine translation, healthcare, speech synthesis and recognition, music understanding and composition), reinforcement learning (with applications to games and real-world problems), game theory and multi-agent systems (with applications to cloud computing, online and mobile advertising), and information retrieval and computational advertising. Most recently, he focuses on AI for science, especially molecular modeling and design, drug discovery and design, biochemistry, etc. He got his PhD degree and Bachelor degree both from Tsinghua University. He is a senior member of ACM and IEEE, and an Adjunct Professor (Ph.D advisor) in the University of Science and Technology of China.
\end{IEEEbiographynophoto}

\begin{IEEEbiographynophoto}
{Tianjun Ke} is a senior undergraduate student at Renmin University of China (RUC). His research interest lies in the intersection of statistics and computer science, including machine learning theory, Bayesian statistics, reinforcement learning, and AI4Science. He has published papers at NeurIPS.
\end{IEEEbiographynophoto}

\begin{IEEEbiographynophoto}
{Rui Yan} is now an associate professor with the Gaoling School of Artificial Intelligence, Renmin University of China. Before, he was a tenure track assistant professor at Peking University. For the past 10+ years, he has been working on Artificial Intelligence (AI) for Natural Language Processing (NLP) and other related research fields such as Data Mining (DM), Information Retrieval (IR) and Machine Learning (ML). He has published papers with conferences such as KDD, SIGIR, WWW, ACL, and AAAI, etc. He has been invited to give tutorial talks at EMNLP, WWW, SIGIR, AAAI, IJCAI. He also serves as an area chair/SPC of KDD, SIGIR, ACL, IJCAI, AAAI, EMNLP, and COLING.
\end{IEEEbiographynophoto}

\end{document}